\documentclass[prb,aps,twocolumn,amsmath,amssymb,floatfix,
superscriptaddress]{revtex4}
\usepackage[dvips]{graphics}
\usepackage{color}
\definecolor{dred}{rgb}{0,0,0.6}
\usepackage{graphicx}  
\usepackage{dcolumn}   
\usepackage{bm}        
\usepackage{amssymb}   
\usepackage{amsmath}
\usepackage{braket}
\usepackage[mathscr]{euscript}
\usepackage{color}
\usepackage{hyperref}
\begin{document}

\title{Interface sensitivity on spin transport through a
  three-terminal graphene nanoribbon}

\author{Sudin Ganguly}
\email{sudin@iitg.ac.in}
\affiliation{Department of Physics, Indian Institute of Technology
  Guwahati, Guwahati-781 039, Assam, India}

\author{Saurabh Basu}
\email{saurabh@iitg.ac.in}

\affiliation{Department of Physics, Indian Institute of Technology
  Guwahati, Guwahati-781 039, Assam, India}

\author{Santanu K. Maiti}

\email{santanu.maiti@isical.ac.in}

\affiliation{Physics and Applied Mathematics Unit, Indian Statistical
  Institute, 203 Barrackpore Trunk Road, Kolkata-700 108, India}


\begin{abstract}
Spin dependent transport in a three-terminal graphene nanoribbon (GNR)
is investigated in presence of Rashba spin-orbit interaction. Such a
three-terminal structure is shown to be highly effective in filtering
electron spins from an unpolarized source simultaneously into two
output leads and thus can be used as an efficient spin filter device
compared to a two-terminal one. The study of sensitivity of the
spin-polarized transmission on the location of the outgoing leads
results in interesting consequences and is explored in details. There
exist certain symmetry relations between the two outgoing leads with
regard to spin-polarized transport, especially when they are connected
to the system in a particular manner. We believe that the prototype
presented here can be realized experimentally and hence the results
can also be verified.
\end{abstract}

\maketitle
\section{\label{sec1}Introduction}
In the last two decades, spintronics has emerged as one of the most
active research fields in condensed matter physics, material science,
and nanotechnology. The main goal of spintronics involves future
power-consuming high operating speed, new forms of information storage
and logic devices~\cite{wolf,santanu-epl-18}. The generation of the
spin-polarized beam is the key factor for the spintronic applications
to be achieved. The spin polarization is generally obtained by a
rotating magnetic field~\cite{pzhang-prl} or by connecting the system
to a ferromagnetic metallic lead~\cite{dutta-das}. However, there are
drawbacks in those methods due to the difficulty in the confinement of
a strong magnetic field in a very small region or due to the
conductivity mismatch between the scattering region and the
ferromagnetic metallic lead~\cite{Schmidt}. Hence it is desirable to
generate spin-polarized current intrinsically~\cite{l-l}, which is
possible in presence of the spin-orbit (SO)
interactions~\cite{qfsun-prb71,santanu-jap-11,qfsun-prb73,hfl,fchi-apl,gong-apl,jap-santanu,santanu-epjb}.

Graphene~\cite{novo} has captured wide attention as a suitable
candidate in the spintronic applications~\cite{neto} due to its
several exciting electronic and transport properties. Some of the them
are the achievement of room-temperature spin transport with long
spin-diffusion lengths (up to $\sim 100\,\mu$m)~\cite{luis,tombros,
  zomer,yang,han}, quasirelativistic band structure~\cite{novo,zhang},
unconventional quantum Hall effect~\cite {novo,zhang,vp}, half
metallicity~\cite{jun,lin} and high carrier
mobility~\cite{du,bolotin}. Moreover, the recent experimental
realization of freestanding graphene nanoribbons
(GNRs)~\cite{meyer,moro} has generated renewed interest in
carbon-based materials with exotic properties. GNRs have also the long
spin-diffusion length, spin relaxation time, and electron spin
coherence time~\cite{yazyev-prb,yazyev-prl,cantele}, hence are
suitable for possible spintronic devices.

Narrow stripes of graphene are called GNR. It can have two types of
geometry along the edges, and they are termed as armchair graphene
nanoribbon (AGNR) and zigzag graphene nanoribbon (ZGNR). Irrespective
of the width, the ZGNRs are always metallic, while the AGNRs are
metallic when the lateral width satisfies the condition $N_y = 3M-1$
($M$ is an integer), else the AGNRs are semiconducting in
nature~\cite{fujita}.

Two kinds of SO couplings can be present in graphene, the intrinsic
and the Rashba SO couplings (SOC)~\cite{km1,km2}. The strength of the
intrinsic SOC is negligibly small in pristine graphene (up to $\sim$
0.01-0.05 meV)~\cite{yao-prb,jc-prb}, while the strength of the Rashba
SOC can be enhanced by growing graphene layer on metallic
substrates. Recently, a Rashba splitting about 225 meV in epitaxial
graphene layers grown on the Ni surface~\cite{dedcov} and a giant
Rashba SOC ($\sim$ 600 meV) from Pb intercalation at the graphene-Ir
surface~\cite{calleja} are noted in experiments. Consequently a
variety of graphene-based spintronic devices have been
proposed~\cite{frank,zeng,kim,jozsa,y-t,bennett,cai,chico,qzhang}, for
example, prediction of spin-valve devices based on graphene
nanoribbons exhibit giant magnetoresistance (GMR)~\cite{kim},
spin-valve experiment on GNR~\cite{jozsa}, study of spin polarization
and giant magnetoresistance in GNR~\cite{y-t}, experiments of GNR as
field-effect transistor~\cite{bennett} and p-n junctions~\cite{cai}
using bottom-up fabrication technique and many
more~\cite{chico,qzhang}. However, most of these studies were based on
two-terminal GNRs, and some non-trivial results are always expected in
multi-terminal bridge systems, as the latter configurations may
exhibit multiple responses in all the outgoing leads
simultaneously. Focusing in that direction, in the present work we are
trying to discuss one such phenomenon, viz, spin-dependent transport
properties in a three-terminal bridge setup.

In general, due to the longitudinal mirror symmetry along the finite
width of a two-terminal ZGNR, only the $y$-component of the
spin-polarized transmission has a non-zero value. The other two
components, namely the $x$ and $z$-components of the spin-polarized
transmission can be generated by making asymmetric square
notch~\cite{qzhang}, introducing adatoms~\cite{sudin-mrx}, disorder
etc. However, without perturbing the central scattering region, it is
also possible to generate all the three components of the
spin-polarized transmission ($P_x$, $P_y$ and $P_z$) with the
help of a three-terminal GNR. Thus a three-terminal structure can be
used as an efficient spin filter device over the two-terminal case. A
few studies have been dedicated to exploration of spin-polarized
transport for three-terminal
GNRs~\cite{antonis,en-jia,hsin-han,jacob,lzhang-jpcm}. These studies
are mostly based on the exploration of electronic and spintronic
properties of different shapes of three-terminal GNR (T-shaped,
fork-shaped, Y-shaped etc.) and also for the rectification and
detection of spin currents. However, we believe that a
deeper look is still needed in order to understand several important
issues which have not been discussed so far. For example, a clear understanding
of the effect of Rashba SO coupling on all three components of the
spin polarized transmission associated with the two outgoing leads in
a three-terminal structure is definitely required for designing efficient spintronic devices. Further, the sensitivity of the
polarization components on the locations of the outgoing leads
attached to the GNR needs a careful scrutiny.

We organize the rest of the paper as follows. In Sec.~\ref{theory}, we present
the model and the theoretical framework for the total transmission and
spin-polarized transmission using the Green's function technique.  In
Sec.~\ref{randd}, we include an elaborate discussion of the results where we
have demonstrated the behavior of the three components of the
spin-polarized transmission, that is, how the spin-polarized
transmission behaves when the location of the outgoing leads are
positioned in several symmetric and asymmetric configurations. We end
with a brief summary of our results. Finally, we conclude stating our
findings in Sec.~\ref{conclusion}.

\section{{\label{theory}}Junction setup and theoretical formulation}
\begin{figure}[h]
\centering
\includegraphics[width=0.45\textwidth]{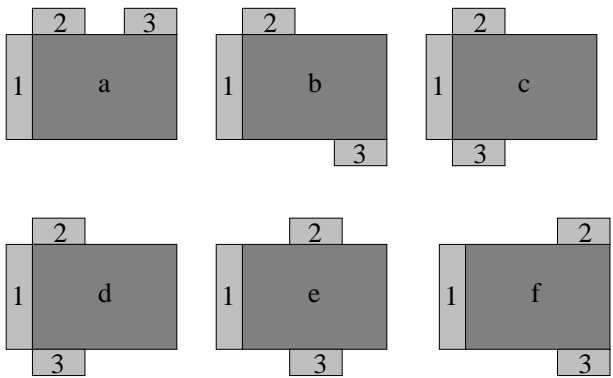}
\caption{Schematic diagram for the three-terminal setup considered in
  the present work. The dark shaded region is the central scattering
  region. The input lead is denoted by lead-1 and the outgoing leads
  are labeled as lead-2 and lead-3. (a-c) lead-3 is moving away from
  lead-2. (d-f) Lead-2 and lead-3 are connected symmetrically to the
  system with respect to lead-1 and are moving away from it.}
\label{setup}
\end{figure}

The schematic diagram is shown in Fig.~\ref{setup} to calculate the
total and spin-polarized transmissions. In all the setups presented in
Fig.~\ref{setup}, the dark shaded region is the scattering region and
the light shaded regions denote the leads attached to it. Lead-1 is
attached to the left side of the central scattering region and lead-2
and lead-3 are attached at either top or bottom sides,
which depends on the configuration of the system. Here, lead-1 is
acting as an input to the system while lead-2 and lead-3 are the
outgoing leads. Through lead-1, unpolarized electrons enter into the
scattering region and in presence of Rashba spin-orbit interaction,
one would expect to observe different kinds of spin species at lead-2
and lead-3 separately. In Fig.~\ref{setup}(a-c), we have fixed the
positions of lead-1 and lead-2 and varied the position of lead-3. In
Fig.~\ref{setup}(d-f), lead-2 and lead-3 are symmetrically connected
to the central scattering region with respect to lead-1.

\begin{figure}[h]
\centering
\includegraphics[width=0.4\textwidth]{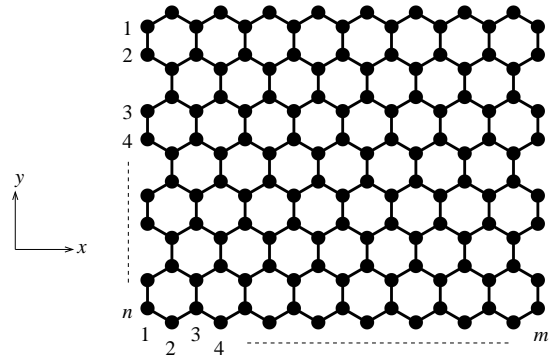}
\caption{Real view of the central scattering region for calculating
  the spintronic properties. Edge passivated scenario is not
    considered here. Leads are not shown here for the better
  understanding to calculate the dimension of the central scattering
  region.}
\label{dim}
\end{figure}
The length and width of the system are measured in the conventional
way as shown in Fig.~\ref{dim}. Along the $x$-direction, the system
has a zigzag shape and along the $y$-direction it is armchair. Hence
the system can be defined as $mZ-nA$. The length and width of the
ribbon can be calculated as in the following,

\begin{equation}
L_x = \frac{\sqrt{3}}{2}(m-1)a,\quad L_y = \left(\frac{3}{2}n-1\right)a
\label{dim_cal}
\end{equation}
where $a=0.142$ nm. Moreover, since we are attaching the outgoing
  leads at the transverse edges of the central scattering region at
  different locations, we have not considered the dangling bonds at the
  edges and therefore any passivated scenario is absent in this work.

The model quantum system is simulated by a tight-binding model, and in
presence of Rashba SO coupling the Hamiltonian reads
as~\cite{km1,km2},

\begin{equation}
H= - t\sum\limits_{\langle ij\rangle}c_i^{\dagger} c_j +
i\alpha\sum\limits_{\langle ij\rangle}c_i^{\dagger} \left(
\vec{\sigma} \times {\bf\hat{d}}_{ij}\right)_z c_j
\label{h2}
\end{equation} 
where $c_i^{\dagger}=\left(c_{i\uparrow}^{\dagger} \quad
c_{i\downarrow}^{\dagger}\right)$. $c_{i\sigma}^{\dagger}$
$(\sigma=\uparrow,\downarrow)$ is the creation operator of an electron
at site $i$ with spin $\sigma$. The first term is the nearest-neighbor
hopping term, with a hopping strength $t$. The second term is the
nearest-neighbor Rashba term which explicitly violates $z\rightarrow
-z$ symmetry. $\vec{\sigma}$ denotes the Pauli spin matrices and
$\alpha$ is the Rashba SO coupling strength. ${\bf\hat{d}}_{ij}$ is
the unit vector that connects the nearest-neighbor sites $i$ and $j$.

The total transmission coefficient, $T_{mn}$, which describes the
total transmission probability of electrons from lead $m$ to lead $n$
can be calculated via~\cite{caroli,Fisher-Lee,dutta},

\begin{equation}
T_{mn} = \rm{Tr}\left[\Gamma_m {\cal G}_R
  \Gamma_n {\cal G}_A\right]
\end{equation}
where ${\cal G}_{R(A)}$ is the retarded (advance) Green's
function. $\Gamma_{m(n)}$ are the coupling matrices representing the
coupling between the central region and the $m(n)$-th lead.

Finally, the spin-polarized transmission coefficient, $P^\alpha_{mn}$,
which describes the spin-polarized transmission of electrons that are
polarized in a particular direction, $\alpha$ from lead $m$ to lead
$n$, can be calculated using~\cite{chang},
\begin{equation}
P^\alpha_{mn} = \rm{Tr}\left[\hat{\sigma}_\alpha\Gamma_m G_R
  \Gamma_n G_A\right]
\label{p-alpha-def}
\end{equation}
where, $\alpha=x,y,z$ and $\sigma$ denote the Pauli matrices.

\section{{\label{randd}}Results and discussion}

We set the hopping term $t=2.7$ eV~{\cite{neto}}. All the
energies are measured in units of $t$. Throughout this paper, we have
fixed the strength of Rashba coupling strength at $\alpha=0.1$. The
dimension of the scattering region in this work is taken as
401Z-60A. By using Eq.~\ref{dim_cal}, the length of the scattering
region is $L_x =49.2\approx 50$ nm. The width is $L_y=12.64$ nm. The
width of lead-1 is same as that of the scattering region, that is,
$W_1=12.64$ nm and has a zigzag shape. The widths of lead-2 and lead-3
($101Z$) are $W_2=W_3=12.3$ nm and have the armchair shape. Thus the
widths of these three leads are close to each other. The widths of the
outgoing leads have been fixed in such a way that they are metallic in
nature. For most of our numerical calculations, we have used
KWANT~\cite{kwant}.

Throughout this work, the strength of the Rashba SO coupling has been
fixed at $\alpha=0.1$ since this value is very close to the
experimentally realized data~{\cite{dedcov}}. The system dimensions have
also been kept same. We have checked the plots shown in this work for
different values of $\alpha$ and also for different system sizes
(shown in the supplemental material), which differ only in
magnitude for the total transmission probability or in the
spin-polarized transmission. But the qualitative results, specifically
the results obtained in Eq.~{\ref{result}} are valid irrespective of the
strength of Rashba SO coupling and the system dimension.

We have essentially studied the behavior of total transmission $T_{mn}$ and all
the three components of the spin-polarized transmission
$P^\alpha_{mn}$ in two different scenarios. First, we have fixed the
positions of leads 1 and 2, and varied the position of lead-3 (see
Fig.~\ref{setup}(a-c)). In the second case, we have attached the leads
2 and 3 symmetrically with respect to lead-1 and varied the positions
of leads 2 and 3 simultaneously away from lead-1 (see
Fig.~\ref{setup}(d-f)).

\begin{figure}[h]
\centering
\includegraphics[height=0.3\textwidth, width=0.4\textwidth]{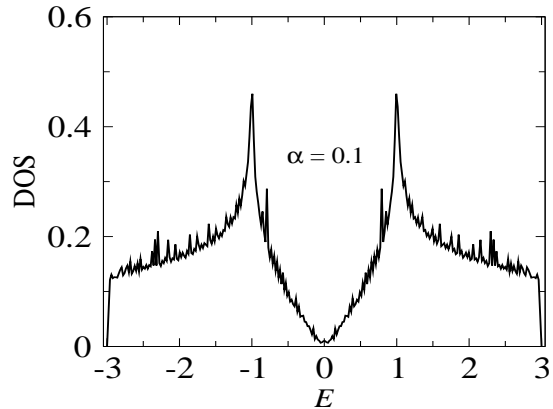}
\caption{(Color online) Density of states (DOS) as a function of the
  Fermi energy.}
\label{dos}
\end{figure}

Before going into the essential results, let us start with the
  variation of density of states (DOS) as a function of the Fermi
  energy which always gives clear picture of the allowed energy zone for electronic transmission. The results are shown in Fig.~{\ref{dos}}. From the spectrum
  we can see that it varies continuously and at $E=0$ there is a sharp
  dip.

Now we focus on the behavior of total transmission probability
as a function of the Fermi energy for the first scenario as mentioned
above.
\begin{figure}[h]
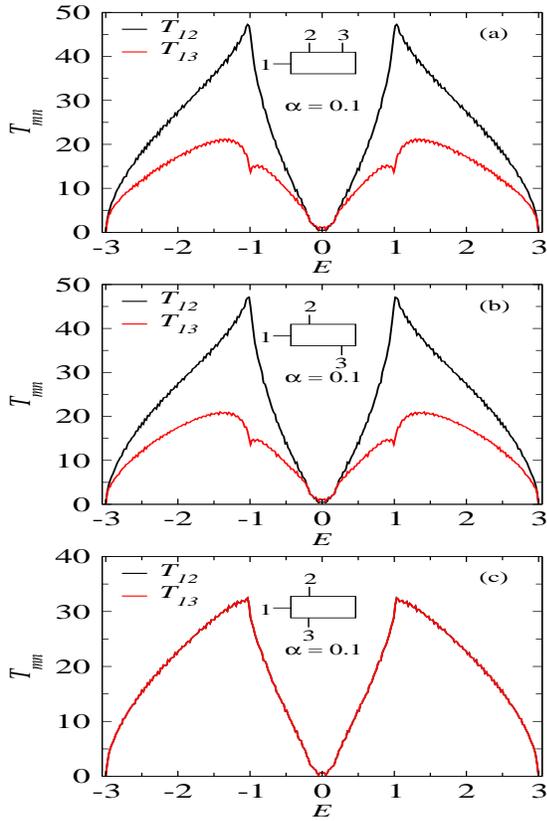

\centering
\includegraphics[height=0.2\textwidth,width=0.4\textwidth]{fig4a.eps}\\
\includegraphics[height=0.2\textwidth,width=0.4\textwidth]{fig4b.eps}\\
\includegraphics[height=0.2\textwidth,width=0.4\textwidth]{fig4c.eps}\\
\caption{(Color online) Total transmission probability $T_{mn}$ as a
  function of the Fermi energy. Lead-3 is attached at the (a) extreme
  top-right (b) extreme bottom-right and (c) extreme bottom-left sides
  of the central scattering region. The brief setups are shown in the
  insets of each spectrum.}
\label{t-asym}
\end{figure}
The results are shown in Fig.~\ref{t-asym}, where three different
cases are considered depending on the specific configurations. The
total transmissions $T_{12}$ and $T_{13}$ are denoted by black and red
colors respectively. When lead-3 is attached at the top-right
(Fig.~\ref{setup}(a)) or at the bottom-right (Fig.~\ref{setup}(b))
side of the central scattering region, both $T_{12}$ and $T_{13}$ show
similar behavior as a function of the Fermi energy. This fact
indicates that electrons do not see any difference whether lead-3 is
connected at the top or bottom of the central scattering
region. However, when lead-3 is attached at the bottom-left side of
the scattering region, $T_{12}$ and $T_{13}$ become exactly same due
to the symmetry of the positions of leads 2 and 3 with respect to
lead-1 (Fig.~\ref{setup}(c))as seen from Fig.~\ref{t-asym} (c). Under
an asymmetric condition, since lead-2 is closer to lead-1 than lead-3,
it is clear that most of the carriers from lead-1 will enter into
lead-2 and remaining ones will enter into lead-3. As a result,
$T_{12}$ will always be higher than $T_{13}$. It is also important to
note that, $T_{12}$ is higher for the asymmetric cases
(Fig.~\ref{t-asym}(a) and Fig.~\ref{t-asym}(b)) than that for the
symmetric case (Fig.~\ref{t-asym}(c)). When lead-2 and lead-3 are
connected symmetrically, probabilities of getting electrons at the two
outgoing leads are same. Along with this fact, due to the effect of
quantum interference among the electronic waves passing through
different arms of the junction, $T_{12}$ ($=T_{13}$) for the symmetric
case becomes always less than that for the asymmetric one. Moreover,
the total transmission spectrum is symmetric about the zero of the
Fermi energy similar to that of the DOS spectrum (shown in Fig.~{\ref{dos}}).

So far, we have studied the total transmission probability for the
three-terminal structure in presence of Rashba SO interaction. Let us
now study the characteristic features of all the three components of
the spin-polarized transmission one by one.

\begin{figure}[h]
\centering
\includegraphics[height=0.35\textwidth,width=0.45\textwidth]{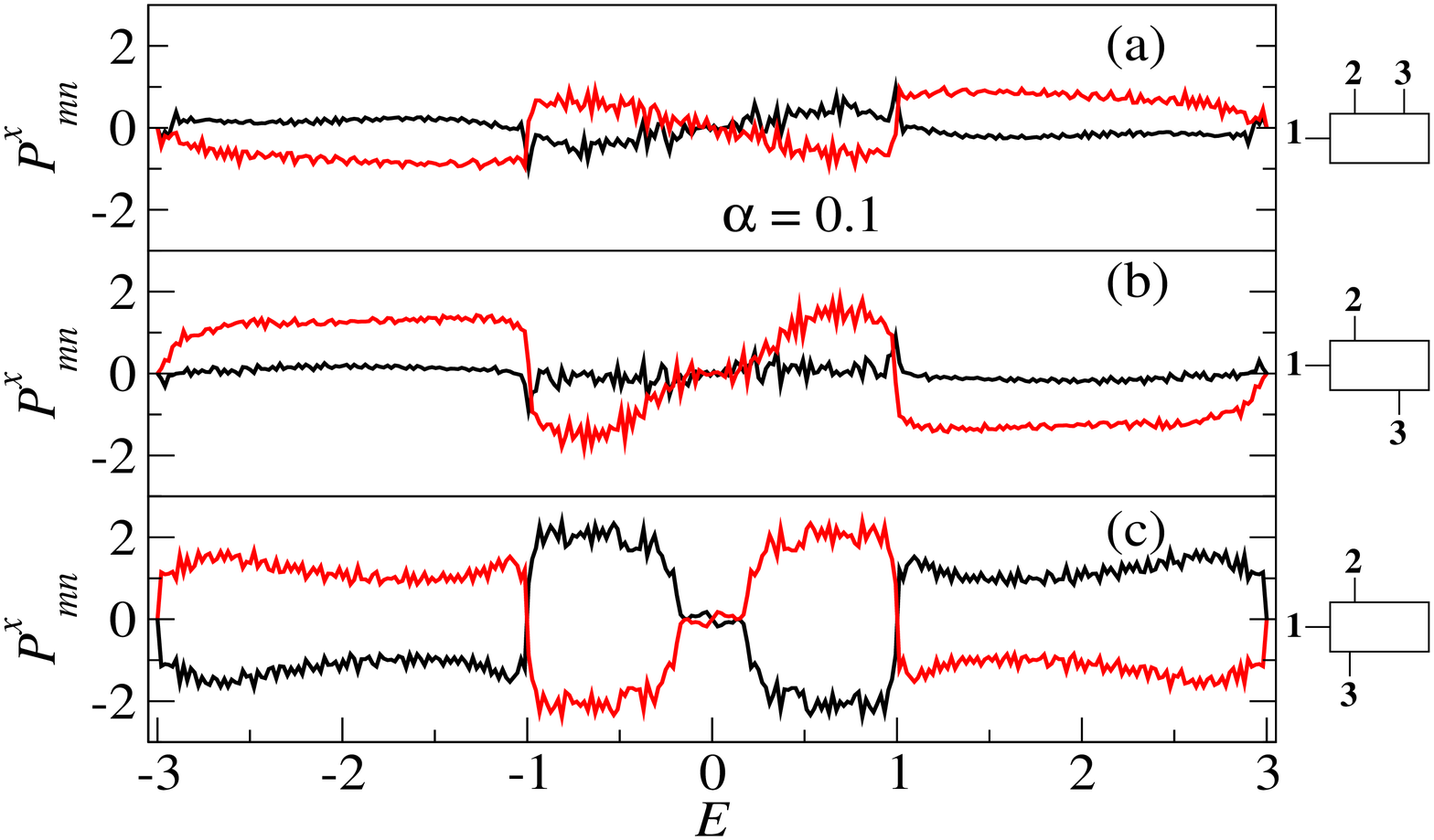}
\caption{(Color online) $x$-component of spin polarization
  coefficient, $P^x_{mn}$, as a function of the Fermi energy in the
  two outgoing leads when lead-3 is coupled at the extreme (a) top-right, (b) bottom-right and (c) bottom-left
  sides of the central scattering region. The black and red colors
  correspond to the results for lead-2 and lead-3 respectively. The
  brief setups are shown at the right side of each plot.}
\label{asym-px}
\end{figure}

Figures~\ref{asym-px}(a-c) show the behavior of the $x$-component of
the spin-polarized transmission, $P^x_{mn}$ as a function of the Fermi
energy. All the plots are antisymmetric about the zero of the Fermi
energy owing to the particle-hole symmetry of the system.
The black color denotes the spin-polarized transmission from lead-1 to
lead-2 and the red color stands for the same from lead-1 to
lead-3. For the first two configurations of the system as shown in
Fig.~\ref{setup}(a) and Fig.~\ref{setup}(b), where lead-3 is attached
at the top-right and bottom-right side of the central scattering
region, $P^x_{12}$ does not change much, but this is not the case for
$P^x_{13}$. In Fig.~\ref{asym-px}(a) and Fig.~\ref{asym-px}(b),
$P^x_{13}$ has opposite signs. For illustration purpose of this
feature, let us look into the region $E>0$. In Fig.~\ref{asym-px}(a),
$P^x_{13}$ has a negative sign, whereas in Fig.~\ref{asym-px}(b), it
is positive. In presence of Rashba spin-orbit interaction, opposite
spins are trying to accumulate in the transverse edges and hence, if
more up spins are available at the top side of the sample than the
down spin or vice versa, there must be a sign difference in $P^x_{13}$
assuming lead-3 is at the top and at the bottom sides of the central
scattering region. Another interesting feature can be inferred from
Fig.~\ref{asym-px}(c) when lead-2 and lead-3 are connected
symmetrically with respect to lead-1. Here $P^x_{12}$ and $P^x_{13}$
are antisymmetric to each other.

The sign difference of the spin-polarized transmission in the two
  outgoing leads has a crucial role in realizing the spin filter
  device. For example, in Fig.~{\ref{asym-px}(a)}, $P^x_{12}$ and
  $P^x_{13}$ have different signs in the energy region
  $E<-1\;(E>1)$. The magnitude and sign reversal of spin
polarization can be explained from the overlap of the up and and down spin bands. The greater asymmetry between
these two spin bands causes higher spin polarization, and when these two bands are completely separated in an energy zone complete polarization can be achieved. Also the
sign of the spin polarization depends on which of, that is, up or down energy band dominates. Both the magnitude and the sign of spin polarization
depend on the SO coupling strength as well as the position of the leads as directly reflected from Eq.~{\ref{p-alpha-def}}. For weak $\alpha$ ($\alpha=0.1$), in the asymmetric case the amplitude will naturally be different as the electronic waves traversing unequal paths before reaching the outgoing leads. Thus a competition ensues between the quantum interference and the SO coupling, and
depending on the dominating one both the sign and the magnitude are
determined. For higher values of $\alpha$, we see that $P_{12}^x$ and $P_{13}^x$ exhibit the same sign for the entire energy window which we verify through our extensive numerical analysis (some of these results are also given in the supplemental material). This argument is also valid for the other two components of spin-polarized transmission. Here it is important to note that for the asymmetric lead-to-conductor configuration, it is quite hard to analyze the sign of polarized spin components
mathematically. However, for the symmetric configuration these sign issues can be clearly explained with mathematical arguments as discussed below in this work.

\begin{figure}[h]
\centering
\includegraphics[height=0.35\textwidth,width=0.4\textwidth]{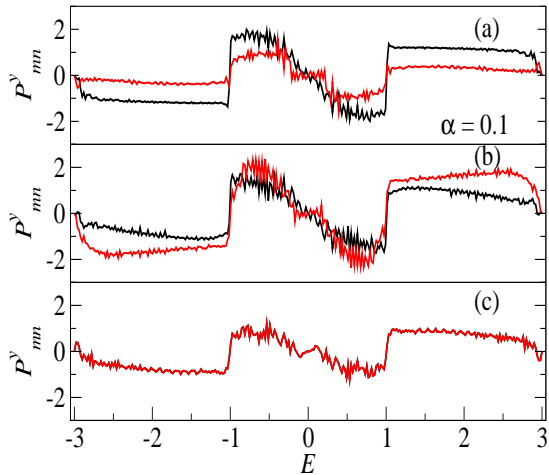}
\caption{(Color online) $y$-component of spin polarization
  coefficient, $P^y_{mn}$ as a function of the Fermi energy where (a),
  (b), and (c) correspond to the identical meaning as given in
  Fig.~\ref{asym-px}. Different colors represent the identical meaning
  as described in Fig.~\ref{asym-px}.}
\label{asym-py}
\end{figure}

The behavior of the $y$-component of the spin-polarized transmission
is shown in Fig.~\ref{asym-py}(a-c) as a function of the Fermi energy.
$P^y_{12}$ is higher than $P^y_{13}$ when lead-2 and lead-3 are
attached on the same side of the system as seen from
Fig.~\ref{asym-py}(a). This indicates that lead-2 is picking up more
$y$-component of spin than lead-3. However, when lead-3 is at the
bottom side of the system, the difference between $P^y_{12}$ and
$P^y_{13}$ become less (Fig.~\ref{asym-py}(b)) and they are exactly
same when lead-2 and lead-3 are connected symmetrically
(Fig.~\ref{asym-py}(c)). Unlike the $x$-component of spin-polarized
transmissions, the $y$-component of the spin-polarized transmissions,
$P^y_{12}$ and $P^y_{13}$ behave similar to total transmission in the
symmetric case.

\begin{figure}[h]
\centering
\includegraphics[height=0.35\textwidth,width=0.4\textwidth]{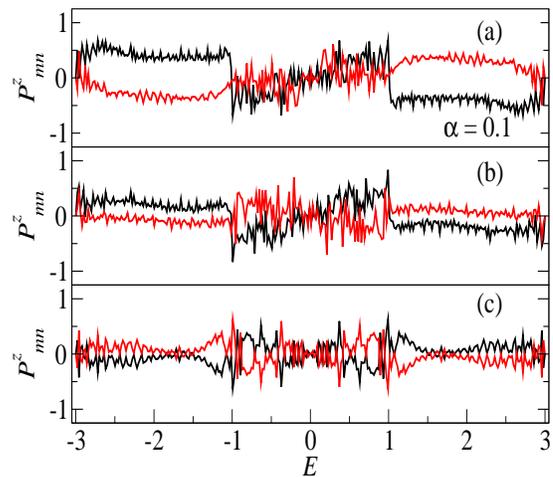}
\caption{(Color online) $z$-component of spin polarization
  coefficient, $P^z_{mn}$ as a function of the Fermi energy where (a),
  (b), and (c) correspond to the identical meaning as given in
  Fig.~\ref{asym-px}. Different colors represent the identical meaning
  as described in Fig.~\ref{asym-px}.}
\label{asym-pz}
\end{figure}

The behavior of the $z$-component of the spin-polarized transmission
is plotted in Fig.~\ref{asym-pz}(a-c) as a function of the Fermi
energy.
In Fig.~\ref{asym-pz}(a), when lead-2 and 3 are on the same side of
the system, $P^z_{12}$ is little higher than $P^z_{13}$. However, when
lead-3 is at the bottom, both the spin-polarized transmissions are
reduced as shown in Fig.~\ref{asym-pz}(b). In Fig.~\ref{asym-pz}(c),
where lead-2 and lead-3 are symmetrically connected, $P^z_{12}$ and
$P^z_{13}$ are antisymmetric to each other.

Comparing the spectra given in Figs.~(\ref{asym-px}-\ref{asym-pz}) we
can see that $P^x_{12}$ is higher when the two output leads are
connected symmetrically with respect to the input lead than the other
two configurations. On the other hand, irrespective of the sign, the
variations of $P^x_{13}$ for the three different configurations are
more or less similar. For the energy region $E<-1$ or $E>1$,
$P^y_{12}$ gets a higher value when two output leads are on the same
side. Again in the same mentioned energy region, the magnitude of
$P^y_{13}$ is in the descending order for the three configurations,
namely when the lead-3 is attached at the bottom-right, bottom-left,
and top-right positions. The $z$-component of the spin-polarized
transmission, $P^z_{12}$ has the highest value when the two output
leads are on the same side and have the lowest value for the
bottom-left configuration. $P^z_{13}$ shows exactly opposite behavior
with respect to $P^z_{12}$.

Now, we shall focus on the symmetric configurations of the two output
leads as shown in Fig.~\ref{setup}(d-f). Here the three different
configurations correspond to the cases when the two output leads are
symmetrically attached at the extreme left (Fig.~\ref{setup}(d)),
middle (Fig.~\ref{setup}(e)) and extreme right (Fig.~\ref{setup}(f))
sides of the central scattering region.
\begin{figure}[h]
\centering
\includegraphics[height=0.35\textwidth,width=0.45\textwidth]{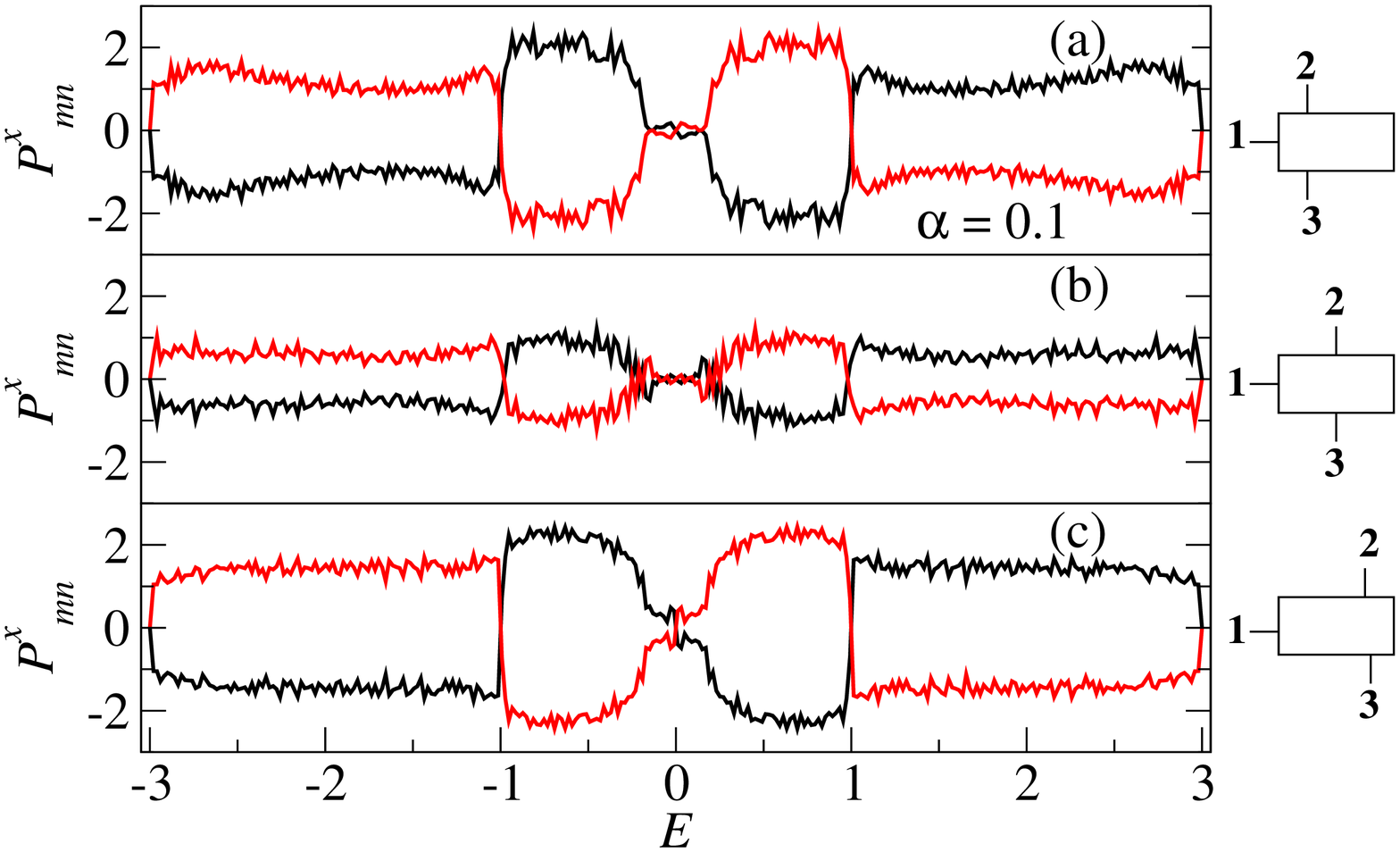}
\caption{(Color online) $x$-component of spin polarization
  coefficient, $P^x_{mn}$ as a function of the Fermi energy for the
  symmetric case. The outgoing leads are connected at the (a) extreme
  left, (b) middle and (c) extreme right sides of the central
  scattering region. Different colors represent the identical meaning
  as described in Fig.~\ref{asym-px}. The brief setups are shown at
  the right side of each plot.}
\label{sym-px}
\end{figure}
In all those setups, the two symmetrically coupled output leads are
positioned away from the input lead and want to study if there is any
effect on the distance between the input and output leads on the
spin-polarized transport properties.
\begin{figure}[h]
\centering
\includegraphics[height=0.35\textwidth,width=0.4\textwidth]{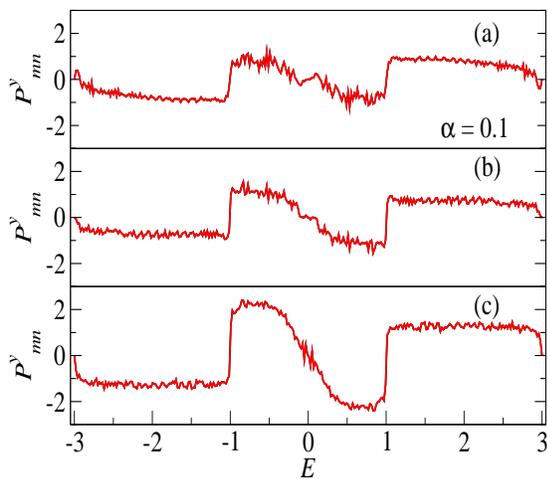}
\caption{(Color online) $y$-component of spin polarization
  coefficient, $P^y_{mn}$ as a function of the Fermi energy where (a),
  (b), and (c) correspond to the identical meaning as given in
  Fig.~\ref{sym-px}. Different colors represent the identical meaning
  as described in Fig.~\ref{asym-px}.}
\label{sym-py}
\end{figure}

\begin{figure}[h]
\centering
\includegraphics[height=0.35\textwidth,width=0.4\textwidth]{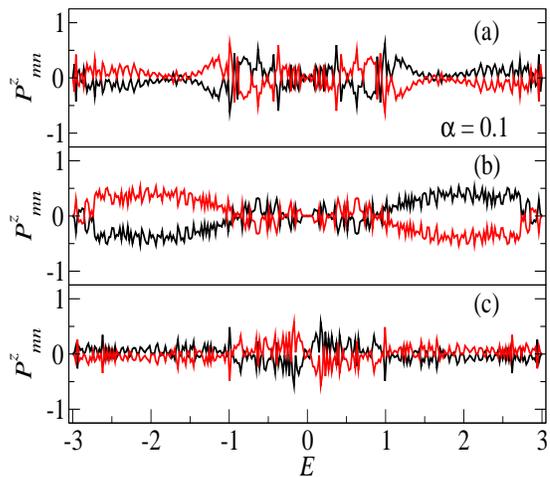}
\caption{(Color online) $z$-component of spin polarization
  coefficient, $P^z_{mn}$ as a function of the Fermi energy where (a),
  (b), and (c) correspond to the identical meaning as given in
  Fig.~\ref{sym-px}. Different colors represent the identical meaning
  as described in Fig.~\ref{asym-px}.}
\label{sym-pz}
\end{figure}

The characteristic features of $P^\alpha_{12}$ and $P^\alpha_{13}$ as
a function of the Fermi energy are shown in Fig.~\ref{sym-px},
Fig.~\ref{sym-py} and Fig.~\ref{sym-pz}. The black lines denote the
results for the output lead-2, while for the other lead (lead-3), the
results are presented by the red lines. From the behavior of the
spin-polarized transmissions, it is observed that the $x$ and
$z$-components of the spin-polarized transmission in two symmetrically
coupled output leads carry
opposite signs as a function of the Fermi energy. In other words,
$P^\alpha_{12}$ and $P^\alpha_{13}$ ($\alpha\in x,z$) are
antisymmetric to each other. On the other hand, the $y$-components in
two symmetrically coupled output leads are exactly same as a function
of the Fermi energy. These features are explained in the following
arguments from the point of view of the structural symmetry of the
system. Moreover, the overall magnitude of the $x$-component is
  lower for the configuration in which the two outgoing leads are
  placed at the middle of the central scattering region than the other
  two configurations. This is entirely due to the quantum interference
  among the electronic waves flowing through the output leads. When
  the two output leads are at either extremes of the central
  scattering region, the constructive interference dominates compared
  to the case where the two outgoing leads are placed at the middle.

The systems described in Fig.~\ref{setup}(d-f) are symmetric with
respect to the reflection $y\rightarrow -y$~\cite{kim-jap}. This
mirror-symmetry of the $y$-axis has an effect on the S-matrix
elements, and by analyzing the S-matrix symmetry Kiselev {\it et
  al.}~\cite{kim-jap} have shown analytically that in a T-shaped
conductor in presence of SO interaction, the transmission amplitudes
for the $x$ and $z$ components have equal magnitude and opposite
phases for symmetrically connected output leads. While, the
$y$-component has identical phase in the output leads with equal
magnitude. Thus we can summarize our observations in a compact way as,
\begin{eqnarray}
\left. 
\begin{aligned}
T_{12}(E) &=& T_{13}(E)\\
P^x_{12}(E) &=& -P^x_{13}(E)\\
 P^y_{12}(E) &=& P^y_{13}(E) \\
P^z_{12}(E) &=& -P^z_{13}(E) 
\end{aligned}
\right\}
\label{result}
\end{eqnarray}

Further, from Fig.~\ref{sym-pz}, it is noted that the $z$-component of
the spin-polarized transmission is much lower than the other two
components (approximately 3 times). Moreover, the values of the $x$
and $y$ components are relatively close to each other.

\section{{\label{conclusion}}Conclusion}
To conclude, in the present work, we have studied spin dependent
transport through a three-terminal GNR in presence of Rashba SO
interaction exploiting the effect of quantum interference among the
electronic waves passing through different arms of the
junction. The three-terminal structure aides in generating all the
  three components of the spin-polarized transmission ($P_x$, $P_y$
  and $P_z$) which was not feasible in a two-terminal GNR and hence it
  can be used as an efficient spin filter device compared to a
  two-terminal one. In addition to that, those spin-polarized
  transmissions can be obtained simultaneously through the two
  outgoing leads in a three-terminal device. Thus, more spin operations are performed simultaneously which would not have been possible for a setup with only one outgoing lead.
 Two different scenarios
have been considered. First we have fixed one outgoing lead (lead-2)
and moved the other outgoing lead (lead-3) away from the other and in
this case, we have an asymmetric geometry. The second scenario is
based on the symmetric configuration of the system, where two leads
are attached symmetrically to the system with respect to the input
lead (lead-1). The $x$ and $z$-components of the spin-polarized
transmission have higher values for the symmetric case, whereas the
total transmission $T$ and the $y$-component of the spin-polarized
transmission are higher corresponding to the case when both the
outgoing leads are on the same side than the other configurations of
the system. Moreover, since the $x$ and $z$-components of the
spin-polarized transmission have the opposite signs in two
symmetrically coupled output leads as a function of the Fermi energy,
the symmetric setup can be used as a switching device.

\acknowledgments

SB thanks Science \& Engineering Research Board, New Delhi, Government
of India, for financial support under the grant F. No:
EMR/2015/001039.

\pagebreak
\widetext
\begin{center}
\textbf{\large Supplemental Materials: Interface sensitivity on spin
  transport through a three-terminal graphene nanoribbon}
\end{center}
\setcounter{equation}{0}
\setcounter{figure}{0}
\setcounter{table}{0}
\setcounter{page}{1}
\makeatletter
\renewcommand{\theequation}{S\arabic{equation}}
\renewcommand{\thefigure}{S\arabic{figure}}
\renewcommand{\bibnumfmt}[1]{[S#1]}
\renewcommand{\citenumfont}[1]{S#1}

The main text discusses the spintronic properties of a three-terminal
graphene nanoribbon (GNR) in presence of Rashba spin-orbit
coupling, where all the results have been computed considering the Rashba coupling strength
$\alpha=0.1$ and the dimension of the central scattering region as
401Z-60A. The width of lead-1 is same as that of the scattering
region, that is, 60A and that of lead-2 and lead-3 are both 101Z. In
this supplementary material we have shown results for density of
states (DOS), total transmission coefficient and the three components
of the spin-polarized transmission ($P_x$, $P_y$ and $P_z$) for
different values of $\alpha$ and for different dimensions of the
\begin{figure}[h]
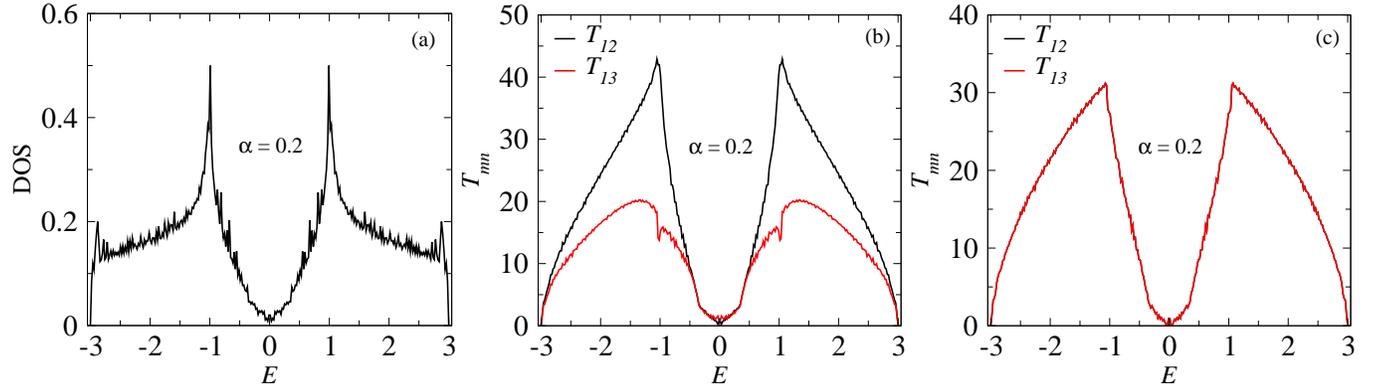

\hfill
\includegraphics[width=0.33\textwidth]{dos_alpha0.2.eps}\hfill
\includegraphics[width=0.33\textwidth]{t_asym150_alpha0.2.eps}\hfill
\includegraphics[width=0.33\textwidth]{t_sym150_alpha0.2.eps}\hfill
\caption{(Color online) (a) Density of states (DOS) as a function of
  the Fermi energy. Transmission coefficients as a function of the
  Fermi energy for (b) asymmetric configuration corresponding to
  Fig.1(a) in the main paper and (c) symmetric configuration
  corresponding to Fig.1(e) in the main paper. The Rashba spin-orbit
  coupling strength is $\alpha=0.2$. The black and red colors
  represent the results for lead-2 and lead-3 respectively.}
\label{1}
\end{figure}
central scattering region. Here the dimension of the scattering region
in Fig.~\ref{1} and Fig.~\ref{2} is taken as 601Z-60A and that in
Fig.~\ref{3} and Fig.~\ref{4} as 801Z-60A. The widths of the leads are
kept same as in the main paper.

The DOS as a function of the Fermi energy is shown in
Fig.~\ref{1}(a) for the Rashba coupling strength $\alpha=0.2$. 
If we
compare the DOS here and the corresponding plot as given in Fig.~3 of
the main paper, we see that both DOS have more or less similar
behavior as a function of $E$, though the strengths of the Rashba
coupling are different (in the main text, $\alpha=0.1$).

The total transmission probability amplitudes in the two outgoing
leads are presented as a function of the Fermi energy for $\alpha=0.2$
in Figs.~\ref{1}(b) and (c). Here we have considered only two
configurations of the system, that is, an asymmetric configuration
(Fig.1(a) in the main paper) and a symmetric configuration (Fig.1(e)
in the main paper) as shown in Figs.~\ref{1}(b) and (c), respectively
in this supplemental material. For the asymmetric case, where two
outgoing leads are on the same side of the central scattering region,
$T_{12}$ is greater than $T_{13}$ (defined in the main text) and this
feature is exactly same with the transmission spectra for $\alpha=0.1$
and also for a different dimension of the central scattering
region. Further, the symmetric configuration shows similar behavior as
discussed in the main paper, that is, $T_{12}$ and $T_{13}$ remain
exactly identical.

\begin{figure}[h]
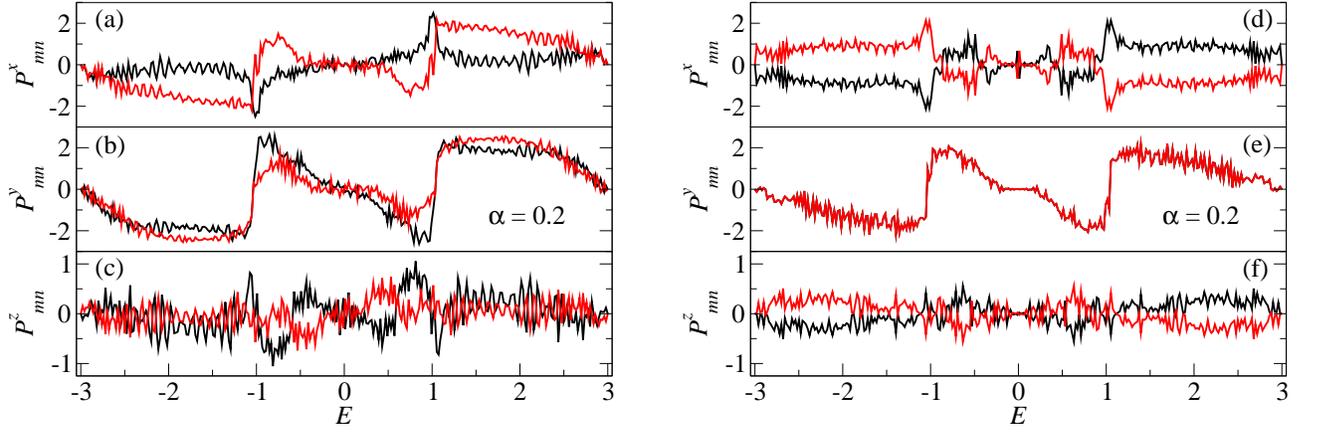

\hfill
\includegraphics[width=0.45\textwidth]{sp_asym150_alpha0.2.eps}\hfill
\includegraphics[width=0.45\textwidth]{sp_sym150_alpha0.2.eps}\hfill
\caption{(Color online) $P_x$, $P_y$ and $P_z$ as a function of the
  Fermi energy for two different configurations, where the left and right columns correspond to the cases as used in Fig.~\ref{1}(b) and Fig.~\ref{1}(c)
  respectively. The Rashba spin-orbit coupling strength $\alpha$ is fixed at 0.2. The black and red colors represent the outputs for lead-2 and lead-3 respectively.}
\label{2}
\end{figure}
The three components of the spin-polarized transmission, $P_x$, $P_y$
and $P_z$ are given as a function of the Fermi energy both for the
asymmetric (results are shown in Fig.~\ref{2}(a-c)) and symmetric cases (results are shown in Fig.~\ref{2}(d-f)). The strength of the Rashba coupling and the
system dimension are same as in Fig.~\ref{1}. The qualitative features
of the spin-polarized transmission spectra are similar to the features
discussed in the main paper. 

The DOS for a higher value of
$\alpha=0.5$ and also corresponding to a higher dimension (601Z-60A)
is shown in Fig.~\ref{3}(a) as a function of the Fermi energy which
has similar behavior as observed in Fig.~\ref{1}(a). The total
transmission probability amplitudes in the two outgoing leads as a
function of the Fermi energy are shown in Fig.~\ref{3}(b) for the
asymmetric case and in Fig.~\ref{3}(c) for the symmetric case. These
plots have similar features as we have discussed earlier for the other
values of Rashba coupling strength and for different dimensions of the
system.

In Fig.~\ref{4}, the variations of $P_x$, $P_y$ and $P_z$ are
presented as a function of the Fermi energy for the same parameters as
used in Fig.~\ref{3}. Owing to the higher value of $\alpha$ (=0.5) ,
\begin{figure}[h]
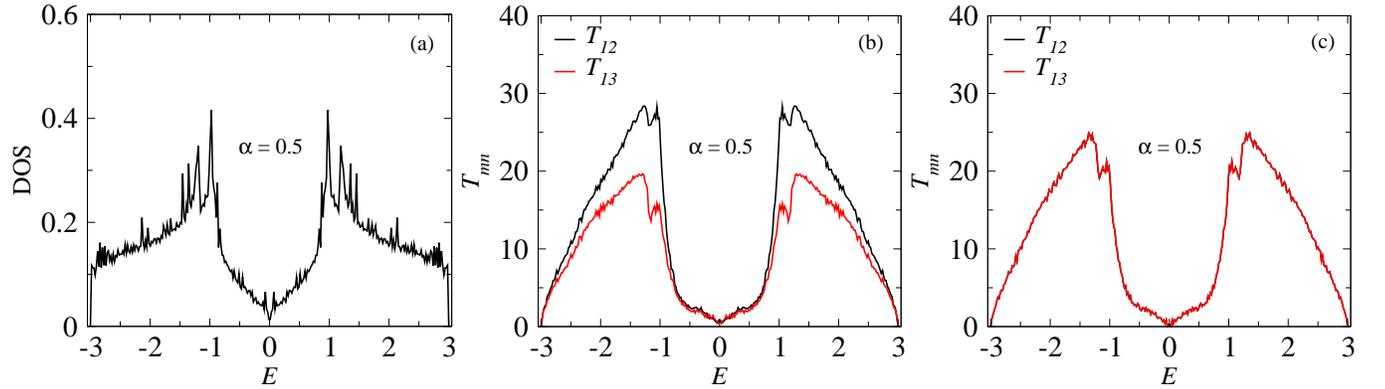

\hfill
\includegraphics[width=0.33\textwidth]{dos_alpha0.5.eps}\hfill
\includegraphics[width=0.33\textwidth]{t_asym200_alpha0.5.eps}\hfill
\includegraphics[width=0.33\textwidth]{t_sym200_alpha0.5.eps}\hfill
\caption{(Color online) (a) Density of states (DOS) as a function of
  the Fermi energy. Transmission coefficients as a function of the
  Fermi energy for (b) asymmetric configuration and (c) symmetric
  configuration as used in Fig.~\ref{1}(b) and Fig.~\ref{1}(c)
  respectively. The Rashba spin-orbit coupling strength is
  $\alpha=0.5$. Different colors represent the identical meaning as
  described in Fig.~\ref{1}.}
\label{3}
\end{figure}
\begin{figure}[h]
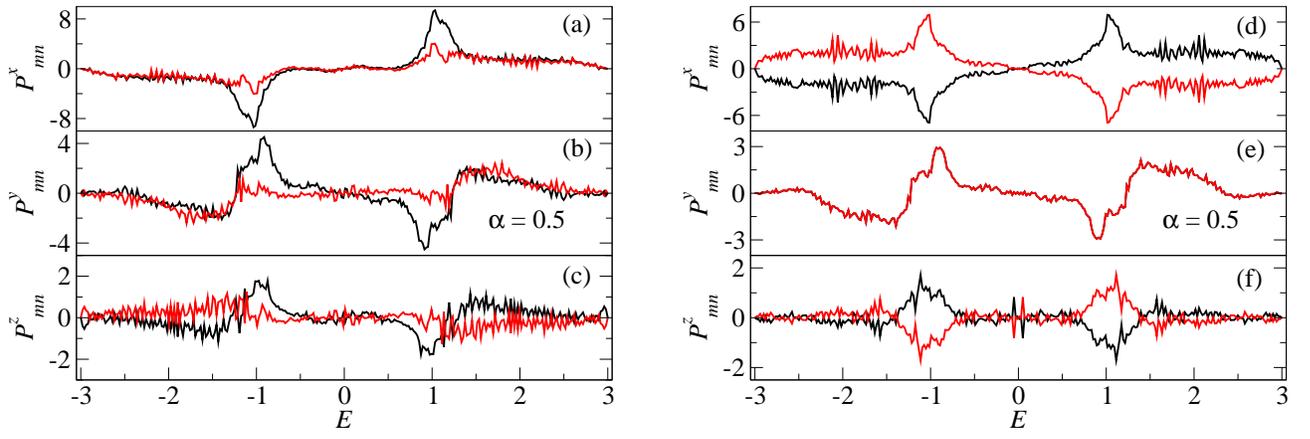

\hfill
\includegraphics[width=0.45\textwidth]{sp_asym200_alpha0.5.eps}\hfill
\includegraphics[width=0.45\textwidth]{sp_sym200_alpha0.5.eps}\hfill
\caption{(Color online) $P_x$, $P_y$ and $P_z$ as a function of the
  Fermi energy for two different configurations, where the left and right columns correspond to the cases as used in Fig.~\ref{1}(b) and Fig.~\ref{1}(c)
  respectively. The Rashba spin-orbit coupling strength $\alpha$ is fixed at 0.5. Different colors represent the identical meaning as
  described in Fig.~\ref{2}.}
\label{4}
\end{figure}the
magnitude of the spin-polarized transmission is greater than the
previous cases. However, the qualitative features remain independent
of the value of the Rashba coupling strength and the system
dimensions. Thus, we can strongly argue that the results presented
here are valid for a wide range of parameter values which prove the
robustness of our analysis. We believe that the studied results will
bring significant impact in analyzing selective spin transmission
through multi-terminal systems comprising different topological
systems.

\end{document}